\documentclass[preprint,showpacs,amsmath,amssymb,aps,prd,nofootinbib]{revtex4}
\usepackage{epsfig}
\usepackage{color}
\def\bea{\begin{eqnarray}}
\def\eea{\end{eqnarray}}

\def\ep{\epsilon}

\renewcommand\epsilon{\varepsilon}

\def\lsim{\mathrel{\raise.3ex\hbox{$<$\kern-.75em\lower1ex\hbox{$\sim$}}} }
\def\gsim{\mathrel{\raise.3ex\hbox{$>$\kern-.75em\lower1ex\hbox{$\sim$}}} }
\begin{document}
\draft
\title{\mbox{}\\[10pt]
Revisiting the Quark-Lepton Complementarity and Triminimal Parametrization of Neutrino Mixing Matrix}

\author{Sin~Kyu~Kang$^1$\footnote{E-mail:
        skkang@seoultech.ac.kr}}

\affiliation{School of Liberal Arts, Seoul National University of Science and Technology, Seoul 139-931, Korea}



\date{\today}

\begin{abstract}
\noindent
We examine how a parametrization of neutrino mixing matrix reflecting quark-lepton complementarity can be probed by considering phase-averaged oscillation probabilities, flavor composition of neutrino fluxes coming from atmospheric and astrophysical neutrinos and lepton flavor violating radiative decays.
We discuss about some distinct features of the parametrization by comparing with the triminimal parametrization of perturbations to tri-bimaximal neutrino mixing matrix.

\end{abstract}
 \maketitle

The enormous progress made in solar, atmospheric and terrestrial neutrino
experiments \cite{PDG} provides us with very robust evidence for the existence of neutrino
oscillations, a new window to physics beyond the standard model.
The current global fits of the neutrino mixing angles are given at the $1(3)\sigma$ level by
\cite{global1}
\begin{eqnarray}
\theta_{12} &=& 34.4\pm 1.0~\left(^{+3.2}_{-2.9}\right)^{\circ} \nonumber \\
\theta_{23} &=& 42.8^{+4.7}_{-2.9}~\left(^{+10.7}_{-7.3}\right)^{\circ}  \label{fit1} \\
\theta_{13} &=& 5.6^{+3.0}_{-2.7}~(\leq 12.5)^{\circ}. \nonumber
\end{eqnarray}
Those results are well consistent with the so-called tri-bimaximal (TB) neutrino
mixing pattern \cite{tribi}
\begin{eqnarray}
U_0=\left(
\begin{array}{ccc}
\frac{2}{\sqrt{6}} & \frac{1}{\sqrt{3}} & 0 \\
\frac{-1}{\sqrt{6}} & \frac{1}{\sqrt{3}} & \frac{1}{\sqrt{2}}  \\
\frac{1}{\sqrt{6}} & \frac{-1}{\sqrt{3}} & \frac{1}{\sqrt{2}}  \\
\end{array}
\right).
\end{eqnarray}
It corresponds to $\sin^2\theta_{12} = 1/3, ~~\sin^2\theta_{23} = 1/2$ and $\sin^2 \theta_{13} = 0$.
Although TB mixing can be achieved by imposing some flavor symmetries, it is widely accepted that
TB is a good zeroth order approximation to reality and there may be deviations from TB in general \cite{dev1}.
With this in mind, it is meaningful to parameterize the lepton mixing matrix in such a way that deviations from TB are manifest.
A useful parametrization of the lepton mixing matrix, so-called ``triminimal" parametrization,  has been proposed such that a mixing angle
in the mixing matrix $\theta_{ij}$ is given by the sum of a zeroth order angle $\theta_{ij}^0$ and a small perturbation
$\epsilon_{ij}$
\cite{trimin}.
A merit of this parametrization is that it leads to simple formulas for neutrino flavor mixing
so that the effects of deviations from the TB mixing could be easily probed
as shown in \cite{trimin}, and this feature is not shared by other parametrizations.

On the other hand, it has been noted that the solar and atmospheric neutrino mixing angles
$\theta_{12}$ and $\theta_{23}$ measured from neutrino oscillation experiments and the quark mixing angles $\theta_{q_{12}}$
and $\theta_{q_{23}}$ reveal a surprising
relation
\bea
\theta_{12}+\theta_{q_{12}} \simeq \theta_{23}+\theta_{q_{23}} \simeq 45^{\circ}, \label{qlc}
\eea
which is satisfied by the experimental results
$\theta_{12}+\theta_{q_{12}}=47.4\pm 1.1\left(^{+3.3}_{-3.0}\right)^{\circ} $  and
$\theta_{23}+\theta_{q_{23}}=45.2^{+4.2}_{-2.9}\left(^{+10.8}_{-7.4}\right)^{\circ} $ to within a
few percent accuracy \cite{global1, PDG}.
This quark-lepton complementarity (QLC) relation (\ref{qlc}) has been interpreted  as an
evidence for certain quark-lepton symmetry or quark-lepton unification as shown in \cite{raidal}.
In the light of the QLC, it is still experimentally allowed for the neutrino mixing matrix
to be composed of a CKM-like matrix and maximal mixing matrices as shown in \cite{skkang2, cabibbo}.
In \cite{skkang2}, it is shown in the framework of supersymmetric standard model that different combination of the mixing matrix leads to different prediction for the branching ratios of lepton flavor violating decays $l_i \rightarrow l_j \gamma$, which makes it possible to discriminate the possible compositions.
Among possible compositions, in this paper, we consider the following parametrization:
\bea
U_{PMNS}=R_{32}\left(\frac{\pi}{4}\right)U^{\dagger}_{CKM}R_{21}\left(\frac{\pi}{4}\right), \label{qlc-ckm}
\eea
where $U_{CKM}$ denotes the CKM mixing matrix.
The reason why we consider this particular parametrization for the QLC relation is that it is well compared and has similar merit to the triminimal parametrization so that we can simply examine if the effects of deviations from the TB mixing can be compatible with the QLC relation or not by investigating a few observables presented by simple formulas.
The parametrization given by Eq.(\ref{qlc-ckm}) can be obtained from the grand unification or quark-lepton symmetry as shown in \cite{skkang2}.
In some unified gauge group such as $SO(10)$, there exist some relations among the Yukawa matrices:
$Y_{\nu}=Y_u=Y_u^T$ and  $Y_e=Y^T_d$, where $Y_{\nu}, Y_{u}, Y_{d}$ and $Y_e$ denote
the Dirac neutrino, up-type quark, down-type quark and charged lepton Yukawa matrices, respectively.
Then, the so-called PMNS neutrino mixing matrix $U_{PMNS}$ is given by $ U_{\rm PMNS} =
V^T_d U_d U_{\rm CKM}^{\dagger}V_M, $ and thus we obtain Eq.(\ref{qlc-ckm}) by taking
$V_d^TU_d=R_{23}(\pi/4)$ and  $V_M=R_{21}(\pi/4)$, where $U_d, V_d$ correspond to the left-handed and right-handed rotation matrices of
down-type quark Yukawa matrix, respectively, and the mixing matrix $V_M$ represents the diagonalizing matrix of
$Y_{\nu}^{diag}V^{\dagger}_0 M_R^{-1}V^{\ast}_0 Y_{\nu}^{diag}$ with a rotation matrix $V_0$ and right-handed heavy Majorana mass matrix $M_R$.
Thus, this parametrization can be used to probe a signal of the grand unification or quark-lepton symmetry.
From the analysis, one can easily prove that this parametrization leads to QLC with an accuracy of order $O(\lambda^2)$.
From now on, we call the parametrization of neutrino mixing matrix given by Eq.(\ref{qlc-ckm}) ``QLC parametrization".

In this short paper, we will examine how the QLC parametrization reflecting a possible hint of the grand unification or quark-lepton symmetry can be probed by considering phase-averaged oscillation probabilities which can be measured from neutrino experiments, flavor composition of neutrino fluxes coming from atmospheric and astrophysical neutrino sources and lepton flavor violating radiative decays.
It is worthwhile to notice that while consideration of the lepton flavor violating
radiative processes can only be applied to the particular model such as the supersymmetric standard model, that of phase-averaged oscillation
probabilities as well as flavor composition of neutrino fluxes are model independent.
We will also discuss about some distinct features of the QLC parametrization by comparing with the triminimal parametrization of perturbations to tri-bimaximal neutrino mixing matrix.
%

Let us first take  $U_{CKM}$ as the Wolfenstein parametrization \cite{ckm1}
as follows:
\bea
U_{CKM}=\left (
\begin{array}{ccc}
1-\frac{1}{2}\lambda^{2}	& \lambda	& A\lambda^{3}(\rho
-i\eta) \\
-\lambda &
1-\frac{1}{2}\lambda^{2}&
A\lambda^{2} \\
A\lambda^{3}(1-\rho-i\eta)	& -A\lambda^{2}& 1
\end{array}
\right ) \; \label{ckm}
\eea
For our numerical calculation, we use the following inputs given by the Particle Data Group \cite{PDG}:
\begin{eqnarray}
\lambda = 0.2257^{+0.0009}_{-0.0010},~~~~A=0.814^{+0.021}_{-0.022}, \nonumber \\
\bar{\rho}=0.135^{+0.031}_{-0.016},~~~~~~\bar{\eta}=0.349^{+0.015}_{-0.017} \label{ckm-fit}
\end{eqnarray}
where $
\bar{\rho}=\rho - \frac{1}{2}\rho \lambda^2 +O(\lambda^4), ~~
\bar{\eta}=\eta-\frac{1}{2}\eta \lambda^2+O(\lambda^4)$.
Inserting Eq.(\ref{ckm}) into Eq.(\ref{qlc-ckm}), we can present the deviations from maximal mixing of the solar and atmospheric mixing angles in powers of $\lambda$: $\theta_{sol}\simeq \pi/4-\lambda$ and $\theta_{atm}\simeq \pi/4-A\lambda^2$.
We also obtain the mixing angle $\theta_{13}$ which is of order $\lambda^3$.

To evaluate the neutrino mixing probabilities for phased-averaged propagation, $P_{\nu_\alpha \leftrightarrow \nu_\beta}$, which is appropriate when the oscillation phase $\Delta m^2 L/4E$ is very large, we need to know $|U_{\alpha i}|^2$ as well discussed in \cite{trimin}.
From Eq.(\ref{qlc-ckm}), we get the matrix form of $|U_{\alpha i}|^2$ which is defined by $\underline{U}_{\alpha i}\equiv |U_{\alpha i}|^2$,
\bea
\underline{U}=&&\frac{1}{4}\left\{ \left(
\begin{array}{ccc}
2 & 2 & 0 \\
1 & 1 & 2 \\
1 & 1 & 2
\end{array} \right)
+\lambda\left(
\begin{array}{rrr}
4 & -4 & 0 \\
-2 & 2 & 0 \\
-2 & 2 & 0
\end{array} \right) \right.  \label{matrix2} \\
&&+ \left. A\lambda^2\left(
\begin{array}{rrr}
0 & 0 & 0\\
2 & 2 & -1 \\
-2 & -2 & 1
\end{array} \right)
+\lambda^3\left(
\begin{array}{ccc}
-2 & 2 & 0 \\
1-S & -1+S & 0 \\
1+S & -1-S & 0
\end{array} \right) \right\},  \nonumber
\eea
where $S=2A+2A\rho $.
In fact, $\underline{U}_{e3}$ is of order of $\lambda^6$, so we have ignored it.
The neutrino mixing probabilities for phased-averaged propagation is given by $P_{\nu_\alpha \leftrightarrow \nu_\beta}=\sum_{i}
\underline{U}_{\alpha i}\underline{U}_{\beta i}$.
Using  Eq.(\ref{matrix2}), we obtain
\bea
P_{\nu_e \leftrightarrow \nu_e} & \simeq & \frac{1}{2}+2\lambda^2 \nonumber \\
P_{\nu_e \leftrightarrow \nu_\mu} & \simeq & \frac{1}{4}-(1-\frac{1}{2}A)\lambda^2 \nonumber \\
P_{\nu_e \leftrightarrow \nu_\tau} & \simeq & \frac{1}{4}-(1+\frac{1}{2}A)\lambda^2  \label {prob} \\
P_{\nu_\mu \leftrightarrow \nu_\mu} & \simeq & \frac{3}{8} + \frac{1}{2}(1-A)\lambda^2 \nonumber \\
P_{\nu_\mu \leftrightarrow \nu_\tau } & \simeq & \frac{3}{8} + \frac{1}{2}\lambda^2 \nonumber \\
P_{\nu_\tau \leftrightarrow \nu_\tau} & \simeq & \frac{3}{8} + \frac{1}{2}(1+A)\lambda^2 \nonumber .
\eea
Here, it is interesting to observe that the only terms proportional to $\lambda^2$ survive in each
$P_{\nu_\alpha \leftrightarrow \nu_\beta}$. We have observed that the contributions at the next next leading order are of order $\lambda^4$.
Imposing the experimental results for $\lambda$ and $A$, we predict the values of $P_{\nu_\alpha \leftrightarrow \nu_\beta}$ corresponding
to the best fit values in Eq.(\ref{ckm-fit}) as follows;
\bea
P_{\nu_e \leftrightarrow \nu_e} &\simeq & 0.6019, ~~~
P_{\nu_e \leftrightarrow \nu_\mu} \simeq  0.2198,  \nonumber \\
P_{\nu_e \leftrightarrow \nu_\tau} &\simeq & 0.1783, ~~~
P_{\nu_\mu \leftrightarrow \nu_\mu} \simeq  0.3797,  \\
P_{\nu_\mu \leftrightarrow \nu_\tau }&\simeq & 0.4005,~~~
P_{\nu_\tau \leftrightarrow \nu_\tau} \simeq  0.4212. \nonumber
\eea

As discussed in \cite{trimin}, it is also interesting to examine how the phase-averaged mixing matrix
in Eq.~(\ref{matrix2}) modifies the flavor composition of the neutrino fluxes.
The most common source for atmospheric and astrophysical neutrinos is thought to be
pion production and decay.
The pion decay chain generates an initial neutrino flux with flavor composition
given approximately~\cite{lipari} by
$\Phi_e^0 : \Phi_\mu^0 : \Phi_\tau^0 = 1 : 2 : 0$ for the neutrino fluxes.
According to Eq.~(\ref{matrix2}), the fluxes $\Phi_\alpha$ arriving at earth
have a flavor ratio of
\begin{small}
\bea
\Phi_e : \Phi_\mu : \Phi_\tau  &=&
1+4A\lambda^2: 1-\frac{1}{2}A\lambda^2: 1-\frac{1}{2}A\lambda^2 \nonumber \\
&\simeq & 1.2 : 1 : 1.
 \label{flux}
\eea
\end{small}
This result shows that $\nu_\mu\leftrightarrow\nu_\tau$ symmetry is kept in the sense that
$ \Phi_\mu/\Phi_\tau = 1$, which is mainly due to the smallness of $U_{e3}$.
The effects of breaking $\nu_\mu\leftrightarrow\nu_\tau$ symmetry appear at order of $\lambda^4$.

Now, let us study the implication of the parametrization given by Eq.(\ref{qlc-ckm})
reflecting quark-lepton unification by considering the lepton flavor violating (LFV)
decays particularly in the context of supersymmetric standard model (SSM).
As is well known, the LFV decays in SSM can be caused by the misalignment of lepton and
slepton mass matrices \cite{masiero} and the branching ratios of the LFV decays depend on
the specific structure of the neutrino Dirac Yukawa matrix $Y_{\nu}$ \cite{skkang2}.
It is well known that the RG running induces off-diagonal terms in the slepton mass
matrix even for the case of universal slepton masses at GUT scale
\footnote{We note that the RG-induced off-diagonal terms in the slepton mass
matrix is more precisely given by \cite{ellis} $
m^2_{\tilde{l}_{ij}}\simeq
-\frac{1}{8\pi^2}(3m_0^2+A_0^2)\left(Y^{\dagger}_{\nu
ik}\log\frac{M_G}{M_{R_k}}Y_{\nu kj}\right).$
But for the sake of simplicity we assume the log term to be universal in our study.
}
\cite{casas}:
\bea
m^2_{\tilde{l}_{ij}}\simeq
-\frac{1}{8\pi^2}(3m_0^2+A_0^2)(Y^{\prime}_{\nu}Y^{\prime\dagger}_{\nu})_{ij}
\log\frac{M_G}{M_X}, \label{slept}
\eea
where $m_0, A_0$ are universal soft scalar mass and soft trilinear $A$ parameter, and
$M_G$ and $M_X$ denote the GUT scale and the characteristic scale of the right-handed neutrinos at which
off-diagonal contributions are decoupled \cite{casas}, respectively.
Here, the Dirac neutrino Yukawa matrix, $Y^{\prime}_{\nu}$, is defined in the basis where the charged lepton Yukawa matrix and
the heavy Majorana mass matrix are real and diagonal, and thus the term $Y^{\prime}_{\nu}Y^{\prime\dagger}_{\nu}$ can be written as
\bea
Y^{\prime}_{\nu} Y^{\prime\dagger}_{\nu}=
R_{23}\left(\frac{\pi}{4}\right) U^{\dagger}_{\rm CKM}(Y^D_{\nu})^2 U_{\rm CKM}R^{\dagger}_{23} \left(\frac{\pi}{4}\right),
\eea
where  $Y^D_{\nu}$ stands for the diagonal form of the Dirac neutrino Yukawa matrix.
For quark-lepton unification, $Y^D_{\nu} = Y^D_u = y_t \mbox{Diag}[\lambda^{8}, \lambda^{4}, 1]$
where $y_t$ is top quark Yukawa coupling \cite{haba}.
Imposing the above form of $Y^D_{\nu}$, we obtain $(Y^{\prime}_{\nu} Y^{\prime\dagger}_{\nu})_{12}\simeq (Y^{\prime}_{\nu} Y^{\prime\dagger}_{\nu})_{13} \simeq \lambda^3$, which leads to  $Br(\mu\rightarrow e \gamma)/Br(\tau\rightarrow e\gamma)\simeq 1$
and it reflects the $\mu-\tau$ symmetry. Also, one can get $(Y^{\prime}_{\nu} Y^{\prime\dagger}_{\nu})_{12})_{23}\simeq 1$,
so that  $Br(\mu\rightarrow e \gamma)/Br(\tau\rightarrow \mu \gamma)\simeq \lambda^6$.
These results indicate that the branching ratio of the LFV decay $\mu(\tau) \rightarrow e \gamma$ is negligibly small
compared with that of $\tau \rightarrow  \mu \gamma$.
%

Now, let us discuss about the implication of the results obtained from the QLC parametrization
by comparing with the triminimal parametrization of perturbations to tri-bimaximal neutrino mixing matrix.
To accommodate the expected deviations from the TB mixing form studied in the literatures \cite{dev1},
the triminimal parametrization of perturbations to tri-bimaximal neutrino mixing matrix has been
proposed \cite{trimin} as follows:
\begin{eqnarray}
U_{TMin}=R_{32}\left(\frac{\pi}{4}\right)U_{\epsilon}(\epsilon_{32};\epsilon_{13},\delta; \epsilon_{21})
         R_{21}\left(\sin^{-1}\frac{1}{\sqrt{3}}\right) \label{trimin}
\end{eqnarray}
where $R_{ij}(\theta)$ describes a rotation in the $ij-$plane through angle $\theta$,
$U_{\delta}=\mbox{diag}(e^{i\delta/2}, 1, e^{-i\delta/2})$ and $U_{\epsilon}=R_{32}(\epsilon_{32})U^{\dagger}_{\delta}R_{13}(\epsilon_{13}(\epsilon_{13})U_{\delta}R_{21}(\epsilon_{21})$.
From the analysis, we obtain the following relations between both parameterizations.
\bea
\sin\ep_{13}e^{-i\delta} & \simeq & \ep_{13}e^{-i\delta} \simeq A\lambda^3 [ 1-\rho+i\eta ] \\
\sin\ep_{32} & \simeq & \ep_{32} \simeq -A\lambda^2 \\
\sin\ep_{21} & \simeq & \ep_{21} \simeq \frac{s}{c}(1-\frac{\lambda}{cs}+\frac{s^2}{c^2}\lambda^2)
\eea
where $s=(\sqrt{2}-1)/\sqrt{6}, c=(\sqrt{2}+1)/\sqrt{6}$.
The first result of the above relations indicates that the QLC parametrization predicts the size of the neutrino mixing angle $\theta_{13}=\epsilon_{13}$ which is at most of order  $\lambda^3$.
The on-going reactor experiments designed to measure $\theta_{13}$ will test whether the QLC parametrization
is ruled out or not.
The precise measurements of the mixing angles $\theta_{23}$ and $\theta_{12}$ would also be useful to probe the QLC parametrization.
 The determination of $\sin^2\theta_{12}$ to $2\%$ level which is comparable to that of the Cabibbo angle ($\simeq 1.4\%$) can be
 achievable in the reactor neutrino experiments as shown in \cite{reactor2} and that of $\sin^22\theta_{23}$
 to $1\%$ is expected to reach in the JPARC-SK experiment \cite{lbl}.
The QLC parametrization is very predictable because the deviations from two maximal mixing angles can be presented in terms of
the well measured parameters in the CKM matrix.
Although the QLC parametrization looks like leading to similar results from the triminimal parametrization, the results presented in Eqs.(\ref{prob},\ref{flux}) show that the parameter $C$ defined in \cite{trimin} is particularly zero in the QLC parametrization, which is
a distinctive feature of the QLC parametrization.
If future experiments confirm our results obtained from the QLC parametrization, it would be difficult to differentiate
between the QLC parametrization reflecting deviations from the bi-maximal mixing \cite{bimax} and the triminimal parametrization reflecting
deviations from the tri-bimaximal mixing.
Confirmation of our results obtained above may also serve as a possible hint of the grand unification or quark-lepton symmetry.

In conclusion, we have examined how the QLC parametrization reflecting a possible hint of the grand unification or quark-lepton symmetry can be probed by considering phase-averaged oscillation probabilities which can be measured from neutrino experiments, flavor composition of neutrino fluxes coming from atmospheric and astrophysical neutrino sources and the ratios of the branching fractions of lepton flavor violating radiative decays.
We have found that those observables are predicted in terms of the well measured parameters of CKM matrix.
We have discussed about some distinct features of the QLC parametrization by comparing with the triminimal parametrization which has been proposed so that the effects of deviations from the tri-bimaximal mixing could be probed.

%

\acknowledgments{ This work is supported by the KRF Grant funded by the Korean Government
(MOEHRD) (KRF-2006-331-C00069).
}



\begin{thebibliography}{99}
\def\plb#1#2#3{Phys.\ Lett.\       {\bf B#1}, (#3) #2}
\def\npb#1#2#3{Nucl.\ Phys.\       {\bf B#1}, (#3) #2}
\def\prd#1#2#3{Phys.\ Rev.\        {\bf D#1}, (#3) #2}
\def\prl#1#2#3{Phys.\ Rev.\ Lett.\ {\bf #1},  (#3) #2}
\def\mpl#1#2#3{Mod.\ Phys.\ Lett.\ {\bf A#1}, (#3) #2}
\def\rep#1#2#3{Phys.\ Rep.\        {\bf #1},  (#3) #2}
\def\sci#1#2#3{Science             {\bf #1},  (#3) #2}
\def\astro#1#2#3{Astrophys.\ J.\   {\bf #1},  (#3) #2}
\def\epj#1#2#3{Eur.\ Phys.\ J.  {\bf C#1},  (#3) #2}
\def\jhep#1#2#3{JHEP               {\bf #1},  (#3) #2}
\def\jpg#1#2#3{J.\ Phys.\        {\bf G#1},  (#3) #2}
\def\ijmp#1#2#3{Int.\ J.\ Mod.\ Phys.\ {\bf #1},  (#3) #2}
\def\ptp#1#2#3{Prog.\ Theor.\ Phys.\ {\bf #1},  (#3) #2}

\bibitem{PDG} K. Nakamura et al. (Particle Data Group), J.\ Phys.\ {\bf G 37}, 075021 (2010).

\bibitem{global1} M. C. Gonzalez-Garcia, M. Maltoni and J. Salvado, JHEP\ {\bf 04}, 056 (2010).

\bibitem{tribi} P. F. Harrison, D. H. Perkins and W. G. Scott, Phys.\ Lett.\ {\bf B 530}, 167 (2002);
                P. F. Harrison and W. G. Scott, Phys.\ Lett.\ {\bf B 535}, 163 (2002);
                Phys.\ Lett.\ {\bf B 557}, 76 (2003);  Z. Z. Xing,
                 Phys.\ Lett.\ {\bf B 533}, 85 (2002); X. He and A. Zee, Phys.\ Lett.\ {\bf B560}, 87 (2003);
                 E. Ma, Phys.\ Rev.\ Lett.\ {\bf 90}, 221802 (2003); C. I. Low and R. R. Volkas, Phy.\ Rev.\ {\bf D 68}, 033007 (2003);
                 G. Altarelli and F. Feruglio, Nucl.\ Phys.\ {\bf B 720}, 64 (2005);
                 S. Chang, S. K. Kang and Kim Siyeon, Phys.\ Lett.\ {\bf B 597}, 78 (2004).

\bibitem{dev1} For an incomplete list see:
Z.~Z.~Xing,
  Phys.\ Lett.\ {\bf B 533} (2002) 85;
  J.~D.~Bjorken, P.~F.~Harrison and W.~G.~Scott,
  Phys.\ Rev.\ {\bf D 74} (2006) 073012;
F.~Plentinger and W.~Rodejohann,
  Phys.\ Lett.\ {\bf B 625}, 264 (2005);
 S.~Antusch and S.~F.~King,
  Phys.\ Lett.\ {\bf B 631}, 42 (2005);
S.~Luo and Z.~Z.~Xing,
  Phys.\ Lett.\ {\bf B 632}, 341 (2006);
S.~K.~Kang, Z.~Z.~Xing and S.~Zhou, Phys.\ Rev.\ {\bf D 73}, 013001 (2006);
N.~Haba, A.~Watanabe and K.~Yoshioka,
  Phys.\ Rev.\ Lett.\  {\bf 97}, 041601 (2006);
M.~Hirsch, E.~Ma, J.~C.~Romao, J.~W.~F.~Valle and A.~Villanova del Moral,
  Phys.\ Rev.\ {\bf D 75}, 053006 (2007);
 X.~G.~He and A.~Zee,
  Phys.\ Lett.\ {\bf B 645}, 427 (2007);
A.~Dighe, S.~Goswami and W.~Rodejohann,
  Phys.\ Rev.\ {\bf D 75}, 073023 (2007);
 M.~Lindner and W.~Rodejohann,
  JHEP {\bf 0705}, 089 (2007);
K.~A.~Hochmuth, S.~T.~Petcov and W.~Rodejohann,
  Phys.\ Lett.\ {\bf B 654}, 177 (2007);
S. F. King, Phys.\ Lett. {\bf B 659}, 244 (2008);
M.~Honda and M.~Tanimoto, Prog.\ Theor.\ Phys.\ {\bf 119}, 583 (2008);
S.~Luo, Phys.\ Rev.\ {\bf D 78}, 016006 (2008);
A.~Datta, arXiv:0807.0420;
A.~Hayakawa, H.~Ishimori, Y.~Shimizu and M.~Tanimoto, Phys.\ Lett.\ {\bf B 680}, 334 (2009);
Y.~Shimizu and R.~Takahashi, Europhys. Lett. {\bf 93}, 61001 (2011);
T.~Araki, J.~Mei and Z.~Z.~Xing, Phys.\ Lett.\ {\bf B 695}, 165 (2011).

\bibitem{trimin} S. Pakvasa, W. Rodejohann and T. J. Weiler, Phys.\ Rev.\ Lett.\ {\bf 100}, 111801 (2008).

%

\bibitem{raidal} M. Raidal,  Phys.\ Rev.\ Lett.\  {\bf 93} (2004) 161801; H. Minakata and A. Yu. Smirnov,
                 Phys.\ Rev.\  D {\bf 70} (2004) 073009; J. Ferrandis and S. Pakvasa,
                 Phys.\ Lett.\ B {\bf 603}, 184 (2004);
                S. K. Kang, C. S. Kim and J. Lee, Phys.\ Lett.\ {\bf B 619}, 129 (2005);
                P. H. Frampton and R. N. Mohapatra,  JHEP {\bf 0501} (2005) 025; 
                K. M. Patel, Phys.\ Lett.\ {\bf B 695}, 225 (2011).

\bibitem{skkang2} K. Cheung, S. K. Kang, C. S. Kim and J. Lee, Phys.\ Rev.\ {\bf D72}, 036003 (2005).

\bibitem{cabibbo}A. Datta, L. Everett, P. Ramond, Phys.\ Lett.\ {\bf B 620}, 42 (2005).


\bibitem{ckm1} L. Wolfenstein, Phys.\ Rev.\ Lett.\ {\bf 51}, 1945 (1983).
%

\bibitem{lipari} P. Lipari, M. Lusignoli and D. Meloni, Phys.\ Rev.\ {\bf D 75}, 123005 (2007).

\bibitem{masiero} F. Borzumati and A. Masiero,
Phys. Rev. Lett. {\bf 57}, 961 (1986).

\bibitem{ellis} J. R. Ellis, J. Hisano, M. Raidal and Y. Shimizu,
Phys.\ Rev.\ {\bf D 66}, 115013 (2002);
A. Masiero, S. K. Vempati and O. Vives,
New J.\ Phys.\ {\bf 6}, 202 (2004).

\bibitem{casas} J. A. Casas and A. Ibarra,
Nucl.\ Phys.\ B {\bf 618}, 171 (2001).

\bibitem{haba} N. Haba and Y. Shimizu, Phys.\ Lett.\ {\bf B 560}, 133 (2003).

\bibitem{reactor2} H. Minakata, H. Nunokawa, W. J. C. Teves and R. Zukanovich Funchal, Phys.\ Rev.\ {\bf D 71}, 013005 (2005).

\bibitem{lbl} H. Minakata, M. Sonoyama and H. Sugiyama, Phys.\ Rev.\ {\bf D 70}, 113012 (2004).

\bibitem{bimax} See for example, V. D. Barger, S. Pakvasa, T. J. Weiler and K. Whisnant, Phys.\ Lett.\ {\bf B 437}, 107 (1998);
G. Altarelli and F. Feruglio, JHEP\ {\bf 9811}, 021 (1998); H. Fritzsch and Z. Z. Xing, Phys.\ Lett.\ {\bf B 440}, 313 (1998);
S. Davidson and S.F. King, Phys.\ Lett.\ {\bf B 445}, 191 (1998);
M. Tanimoto, Phys.\ Rev.\ {\bf D 59}, 017304 (1998);
Y. Nomura and T. Yanagida, Phys.\ Rev.\ {\bf D 59}, 017303 (1998);
R. N. Mohapatra and S. Nussinov, Phys.\ Rev.\ {\bf D 60}, 013002 (1999);
S. K. Kang and C. S. Kim, Phys.\ Rev.\ {\bf D 59}, 091302 (1999).
H. Fritzsch and
Z. Z. Xing, Prog. Part. Nucl. Phys. {\bf 45}, 1 (2000) and
references therein.
\end{thebibliography}
\end{document}